%% file: main.tex
\documentclass[acmlarge,screen]{acmart}
\AtBeginDocument{%
  \providecommand\BibTeX{{%
    \normalfont B\kern-0.5em{\scshape i\kern-0.25em b}\kern-0.8em\TeX}}}

\setcopyright{acmcopyright}
\copyrightyear{2021}
\acmYear{2021}
\acmDOI{10.1145/1122445.1122456}

\acmConference[IMWUT/Ubicomp' 21]{Proceedings of the ACM on Interactive, Mobile, Wearable and Ubiquitous Technologies}
\acmBooktitle{Proceedings of the ACM on Interactive, Mobile, Wearable and Ubiquitous Technologies}
\acmPrice{15.00}
\acmISBN{978-1-4503-XXXX-X/18/06}

\usepackage{graphicx}
\usepackage{xspace}
\usepackage{mdframed}
\usepackage{enumitem}

\newcommand*{\projectname}{SplitSR\xspace}
\newcommand*{\projectnameblock}{SplitSRBlock\xspace}
\newcommand*{\appname}{ZoomSR\xspace}
\setcopyright{acmcopyright}
\copyrightyear{2021}
\acmYear{2021}
\acmDOI{10.1145/1122445.1122456}
\acmJournal{IMWUT}
\acmVolume{0}
\acmNumber{0}
\acmArticle{0}
\acmMonth{3}
\begin{document}
\newpage

\title{\projectname: An End-to-End Approach to Super-Resolution on Mobile Devices}
\input{sections/authors}
\input{sections/abstract}

\ccsdesc[500]{Human-centered computing~Ubiquitous and mobile computing systems and tools}
\ccsdesc[500]{Human-centered computing~Interactive systems and tools}
\ccsdesc[500]{Computing methodologies~Computer vision tasks}

\keywords{mobile computing, image super-resolution, edge computing}

\maketitle

\input{sections/introduction}
\input{sections/related_work}
\input{sections/method}
\input{sections/experiments}
\input{sections/user_study}

\input{sections/conclusion}

\bibliographystyle{ACM-Reference-Format}
\bibliography{SR}

\newpage

\end{document}

%% file: sections/authors.tex
\author{Xin Liu}
\email{xliu0@cs.washington.edu}
\affiliation{%
  \institution{University of Washington}
  \city{Seattle}
  \state{WA}
  \country{USA}
}

\author{Yuang Li}
\affiliation{%
  \institution{BUPT \& University of Washington}
  \city{Seattle}
  \state{WA}
  \country{USA}
}

\author{Josh Fromm}
\affiliation{%
  \institution{University of Washington \& OctoML}
  \city{Seattle}
  \state{WA}
  \country{USA}
}

\author{Yuntao Wang}
\affiliation{%
  \institution{Tsinghua Univiersity \& University of Washington}
  \city{Seattle}
  \state{WA}
  \country{USA}
}

\author{Ziheng Jiang}
\affiliation{%
  \institution{University of Washington \& OctoML}
  \city{Seattle}
  \state{WA}
  \country{USA}
}

\author{Alex Mariakakis}
\email{mariakakis@cs.toronto.edu}
\affiliation{%
  \institution{University of Toronto}
  \city{Toronto}
  \country{Canada}
}

\author{Shwetak Patel}
\email{shwetak@cs.washington.edu}
\affiliation{%
  \institution{University of Washington}
  \city{Seattle}
  \state{WA}
  \country{USA}
}

%% file: sections/abstract.tex
\begin{abstract}
Super-resolution (SR) is a coveted image processing technique for mobile apps ranging from the basic camera apps to mobile health. Existing SR algorithms rely on deep learning models with significant memory requirements, so they have yet to be deployed on mobile devices and instead operate in the cloud to achieve feasible inference time. This shortcoming prevents existing SR methods from being used in applications that require near real-time latency. In this work, we demonstrate state-of-the-art latency and accuracy for on-device super-resolution using a novel hybrid architecture called \projectname and a novel lightweight residual block called \projectnameblock. The \projectnameblock supports channel-splitting, allowing the residual blocks to retain spatial information while reducing the computation in the channel dimension. \projectname has a hybrid design consisting of standard convolutional blocks and lightweight residual blocks, allowing people to tune \projectname for their computational budget. We evaluate our system on a low-end ARM CPU, demonstrating both higher accuracy and up to 5$\times$ faster inference than previous approaches. We then deploy our model onto a smartphone in an app called \appname to demonstrate the first-ever instance of on-device, deep learning-based SR. We conducted a user study with 15 participants to have them assess the perceived quality of images that were post-processed by \projectname. Relative to bilinear interpolation --- the existing standard for on-device SR --- participants showed a statistically significant preference when looking at both images (Z=-9.270, p<0.01) and text (Z=-6.486, p<0.01).
\end{abstract}

%% file: sections/introduction.tex
\section{Introduction}
\label{sec:intro}
Image super-resolution (SR) is the task of reconstructing a high-resolution image from a low-resolution input. One domain where SR could have major impact is human-computer interaction. For example, most smartphones rely on digital zoom because they do not have an optical zoom lens, so on-device SR can provide high-resolution images when users examine static images~\cite{georgis2013super}. Low-latency, on-device SR can also improve the quality of feedback during image capture as users adjust the camera's position and zoom factor. When users take photographs of documents, they can be confident that they will be able to read their documents later on since on-device SR can make individual characters more discernible~\cite{datsenko2007example,pandey2018binary}. When users want to send photographs to others via computer-mediated communication, on-device SR can reduce data transmission requirements by allowing users to send low-resolution images that are later upsampled~\cite{dasaristreaming}.

Another domain where on-device SR would have significant impact is healthcare~\cite{greenspan2009super, robinson2017new, trinh2014novel}, especially in the developing world where lower-end smartphones are more common. Plug-and-play smartphone accessories like portable ultrasound probes enable medical imaging, with the smartphone serving as a computation and communication hub~\cite{becker2016use}. Researchers have also developed smartphone accessories that augment the capabilities of the built-in camera for health screening~\cite{sung2017open,geng2017recent}, including enclosed optical components for identifying refractive errors~\cite{Pamplona2010} and cataracts~\cite{Pamplona2011}. Furthermore, mobile health researchers have begun to utilize unmodified smartphone cameras for health screening. For example, \citet{Wadhawan2011} created a smartphone app that analyzes images of skin lesions to classify them as benign or malignant, while \citet{Mariakakis2016} proposed an app that analyzes short videos of pupil dilation for identifying traumatic brain injuries. Across all of these healthcare applications, low-latency on-device SR would ensure that small details are not overlooked during image analysis and not affect the user experience. By keeping computation on the smartphone itself and bypassing the need for data upload, connectivity in rural regions becomes less critical and sensitive patient information can be preserved.

Deep convolutional neural networks (CNNs) have yielded state-of-the-art accuracy for SR~\cite{yang2019deep, redmon2016you, krizhevsky2012imagenet}, but achieving this performance has required network architectures with many layers and a large spatial dimension across layers. As a result, these models usually have 1--5 million parameters, making them infeasible to directly deploy on resource-constrained platforms~\cite{yang2019deep, hui2018fast, zhang2018image}. Therefore, such models are often deployed on cloud-based servers, which incurs latency penalties for data upload and result download. The inference time of the models themselves are also not trivial, falling short of the requirements for on-device scenarios~\cite{lee2019mobisr}. Low-latency on-device SR would change this paradigm, eliminating unnecessary data transmission and power consumption during real-time applications.

To overcome these limitations, we propose a novel architecture for on-device SR called \projectname. Our key insight is that a hybrid design consisting of standard convolutions and lightweight residual blocks can be used to maximize accuracy while satisfying a restricted computational budget.
We apply this insight to extend RCAN~\cite{zhang2018image}, a state-of-the-art SR model, so that it can be run on a mobile device. To push the limits of \projectname even further, we also introduce a novel lightweight residual block called the \projectnameblock. This block leverages channel-splitting within standard 2D convolutions to retain spatial information and reduce computational cost. 
We study various configurations of our \projectname architecture and \projectnameblock to see how hyperparameters impact both inference latency and accuracy. We also leverage a modern deep learning compiler called TVM~\cite{chen2018tvm} to implement the proposed system and generate highly efficient machine code that accelerates inference speed on our target embedded device. In doing so, we demonstrate that \projectname achieves approximately 5$\times$ speedup and superior accuracy compared to a state-of-the-art on-device SR model. We then deployed \projectname in an Android app called \appname to see if users could notice the difference in latency and image quality between \projectname and bilinear interpolation, the latter being the existing standard for on-device SR. Through a 15-person user study, we find that the participants preferred \projectname over bilinear interpolation for both images of objects (Z=-9.270, p<0.01) and text (Z=-6.486, p<0.01). In summary, our contributions include: 
\begin{itemize}
  \item A hybrid architecture called \projectname that combines standard convolutional blocks and lightweight residual blocks to enable on-device SR within a constrained computational budget,
  \item A lightweight residual block called a \projectnameblock that applies channel-splitting to standard 2D convolutions in order to accelerate computation and retrain high-quality spatial transformation,
  \item A comprehensive analysis that evaluates the benefits of our \projectnameblock and our \projectname model against state-of-the-art alternatives,
  \item \appname, the first-ever smartphone app to demonstrate on-device, deep learning-based SR, and
  \item A user study with \appname demonstrating that people prefer the image quality engendered by \projectname rather than bilinear interpolation. 
\end{itemize}

%% file: sections/related_work.tex
\section{Related Work}
\label{sec:related}
In this section, we provide an overview of single-image SR solutions, including those designed for on-device systems. We then describe how \projectname improves upon its predecessors.

\subsection{Super Resolution}
\label{sec:related-sr}
Historically, researchers have explored myriad methods for attacking the super-resolution task --- ranging from example-based strategies to sparse-coding-based methods~\cite{yang2019deep, yang2010image, zeyde2010single, timofte2013anchored}. However, convolutional neural networks (CNNs) have quickly overtaken these approaches given their ability to extract a high-level representation of image data. Recent studies have demonstrated the superiority of deep learning compared to alternative methods when it comes to state-of-the-art SR tasks~\cite{dai2019second, zhang2018image}. One of the earliest examples of deep learning for SR is by \citet{dong2015image}. \citet{kim2016accurate} later improved upon Dong et al.'s work with a much deeper convolutional neural network ($\geq$16 layers), using residual learning to overcome the challenges of so many layers. One of the major drawbacks of these works is that they require interpolating the low-resolution images to high-resolution inputs before they are fed into the network, which adds computational complexity and potentially loses significant details due to smoothing effects. By training a separate upsampler module at the tail of the network, \citet{dong2016accelerating} introduced a faster and more accurate network. The idea of using an independent upsampling block has become a common choice for many modern SR architectures.

Numerous enhancements have been made to further improve SR accuracy. 
\citet{zhang2018image} added an attention module to create a residual channel attention network (RCAN). To the best of our knowledge, RCAN's performance is currently rivaled by \citet{dai2019second}'s second-order attention network (SAN) and \citet{liu2019hierarchical}'s hierarchical back projection network (HBPN). However, these approaches incur significant computational costs, with SAN using self-ensembling across multiple SR models and HBPN requiring both RGB and YUV images as input. \citet{sun2020learned} recently proposed a model called a content adaptive resampler (CAR), which improves the performance of a regular SR network by jointly training it with a model that downsamples high-resolution images. Like SAN, CAR incurs high computational costs since it requires a joint architecture, and CAR is also limited by the fact that it requires high-resolution images for inference. These models must be significantly downsized to accommodate the limited computational resources of mobile devices. However, reducing model complexity often leads to decreased accuracy, especially when the downsizing is done na\"ively.
In this work, we strive to create an accurate SR model that is efficient enough to run on a mobile device. \projectname is a hybrid architecture that is intentionally designed to be tuned for a particular computational budget. For tighter budgets, \projectname replaces standard convolutional blocks with lightweight residual blocks in a unique way to prioritize latency over accuracy. We apply this concept to the RCAN~\cite{zhang2018image} architecture since it is more commonly used in the literature.

\subsection{On-Device Super Resolution}
\label{sec:related-ondevice}
Previous work has explored several methods for accelerating SR inference with custom hardware. For example, \citet{kim2018real} demonstrated real-time SR on a highly optimized FPGA-based system. Although they attained a real-time performance from full HD (FHD) to 4K ultra HD (4K UHD), the algorithm was deployed on bulky and dedicated hardware that is not flexible to run other SR models. \citet{dasaristreaming} introduced a system for streaming 360$^\circ$ videos using SR. Although their work focused on network latency, they did not present results for how their approach would work on common 2D image datasets. 

The most similar work to our own is MobiSR by \citet{lee2019mobisr}. MobiSR is an on-device SR system that runs on heterogeneous mobile processors, integrating the CPU, GPU, and digital signal processor (DSP). MobiSR achieves significant speedup compared to prior work, but the authors of that work were unable to implement their system on a mobile device because of the multi-processor load-balancing that was required. \projectname achieves state-of-the-art mobile SR performance in terms of both speed and accuracy using a single neural network and a single CPU, making it accessible to many modern smartphones. In this work, we deploy \projectname in an Android app called \appname and evaluate whether users notice the difference between images processed by \projectname versus bilinear interpolation.

\subsection{Lightweight Residual Blocks}
\label{sec:related-residual}
Standard convolutional layers are the fundamental building blocks for many computer vision applications. When many convolutional layers are stacked together to form very deep networks, residual connections can be used to facilitate gradient flow through the network during training. At a certain point, many networks become too computationally intensive for on-device systems, especially when the channel dimension of the feature maps is large. Assuming the input and output channels of a given convolutional layer are both $N$, the computational complexity of a standard convolutional layer is $O(N^{2})$. In this section, we cover recent optimizations of convolutional layers that attempt to reduce convolutional computational complexity while maintaining high accuracy.

\subsubsection{BottleneckBlock \& ShuffleBlock}
\begin{figure*}[t!]
  \includegraphics[width=\textwidth]{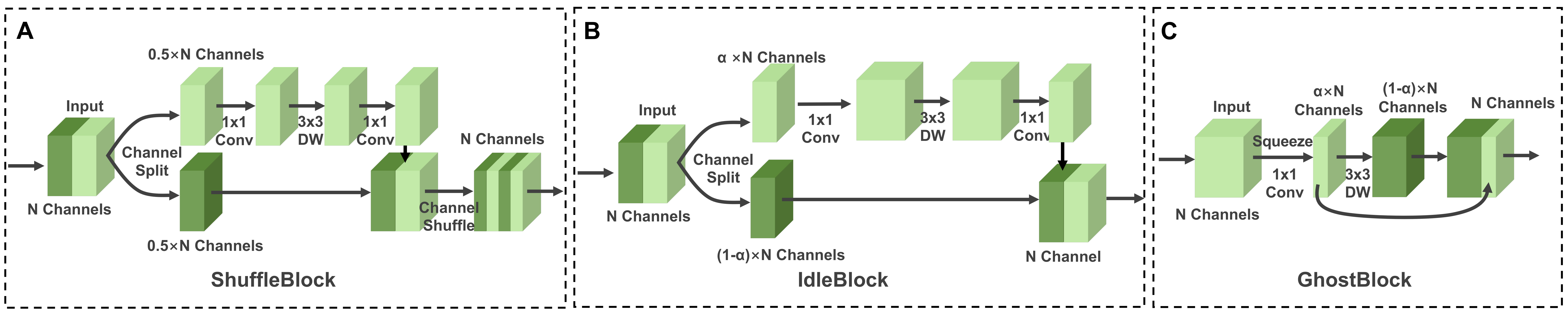}
  \caption{Schematics showing several lightweight residual blocks from prior work: (A) ShuffleBlock~\cite{ma2018shufflenet}, (B) IdleBlock~\cite{xu2019hybrid}, and (C) GhostBlock~\cite{han2019ghostnet}.}
  \label{fig:compare_blocks}
\end{figure*}

\citet{howard2017mobilenets} proposed the use of depthwise separable convolutions to stack residual blocks and achieve a strong balance between accuracy and latency. These operations, also known as BottleneckBlocks, consist of two parts: a 3$\times$3 depthwise convolution and by a 1$\times$1 pointwise convolution. The depthwise convolution is used to expand features and makes use of single convolutional filters at every input channel to perform lightweight filtering, while the pointwise convolution constructs new features by calculating linear combinations of input channels. ShuffleBlock, introduced by \citet{ma2018shufflenet}, is an advanced version of the BottleneckBlock with several innovations (Figure~\ref{fig:compare_blocks}A). First, ShuffleBlocks use channel-splitting to evenly split the input channels into branches. Only one branch is used for computation, where each channel is processed through two 1$\times$1 pointwise convolutions and one 3$\times$3 depthwise convolution. The branches are then concatenated, after which the channels are shuffled to allow information to be communicated between the two branches. 

\subsubsection{Inverted Residual Block \& IdleBlock}
Inverted residual blocks are designed to process significantly more spatial information during the depthwise convolution. To do this, these blocks add an additional pointwise layer at the beginning to expand low-dimensional information to a high-dimensional representation. After going through the depthwise convolution, these blocks use another pointwise layer to project the high-dimensional representation to the original low-dimensional space. \citet{sandler2018mobilenetv2}, who first proposed these blocks as part of their MobileNetV2 work, gained considerable performance improvements performance on the ImageNet dataset. Recently, \citet{xu2019hybrid} proposed the idea of combining an inverted residual block with a ShuffleBlock (without channel-shuffling) in order to prune superfluous connections. This construct is called an IdleBlock (Figure~\ref{fig:compare_blocks}B. 
\citet{xu2019hybrid} also studied the use of IdleBlock in various hybrid configurations using two different type of lightweight residual block for image classification and achieved the state-of-the-art accuracy against other mobile-specific neural networks.

\subsubsection{GhostBlock}
CNNs often contain redundancy in their feature maps. Calling these redundancies ghost features, \citet{han2019ghostnet} recognized this inefficiency as an opportunity for optimization. They proposed the notion of a GhostBlock (Figure~\ref{fig:compare_blocks}C) for generating redundant feature maps with cheap linear operations rather than ordinary convolutions. Within a GhostBlock, an ordinary convolution kernel is applied to generate a few intrinsic feature maps. Linear operations are used to generate redundant feature maps, which are then stacked with the intrinsic feature maps to produce the block's output. A pointwise convolution is used in the GhostBlock's first layer, depthwise convolutions are used for cheap linear operations.  

~\\
We draw inspiration from this prior work to propose the \projectnameblock, which improves performance by applying channel-splitting to standard 2D convolutions. We compare the \projectnameblock against the aforementioned lightweight residual blocks in our \projectname architecture, making us the first to evaluate these blocks in the context of image SR.

%% file: sections/method.tex
\section{Method}
\label{sec:method}

In the following sections, we first describe the design of a novel lightweight residual block called the \projectnameblock. We then describe how such blocks can be used in a novel hybrid architecture called \projectname, which mixes standard convolutional blocks with lightweight residual blocks to enable on-device SR.

\subsection{\projectnameblock Design}
Channel-wise attention has been used to help SR networks like RCAN~\cite{zhang2018image} focus on informative features (i.e., channels) to boost accuracy. Because some channels are generally more important than others, the success of channel-wise attention has shown that not every channel is essential to generating the final high-resolution output. Moreover, channel-splitting can act like channel-attention to provide more complex weights on certain channels through back-propagation. Therefore, we propose a lightweight residual block design that uses channel-splitting to reduce computation, while maintaining high-quality spatial transformation from standard 2D convolutions to achieve high accuracy. The structure of \projectnameblock is inspired by that of previously proposed lightweight residual blocks (Figure~\ref{fig:compare_blocks}), which can be summarized in three stages. First, a tensor with $N$ channels is split along the channel axis into two branches according to parameter $\alpha$. Second, the branch with $\alpha \times N$ channels is used to perform either depthwise or pointwise convolutions while the other branch is left idle. When $\alpha$=0.5, for instance, one branch uses half of the channels to conduct computation, and the other branch passes the remaining channels through the network. Finally, the two branches are concatenated to construct the block's output such that it has the same number of channels as the input. As $\alpha$ decreases, more layers can be added while keeping the same number of total network parameters. This structure is both computationally efficient and able to achieve high accuracy in SR because it is able to learn enough information about the image using only a subset of the input channels and backpropagation.

\begin{figure*}[t!]
  \includegraphics[width=\textwidth]{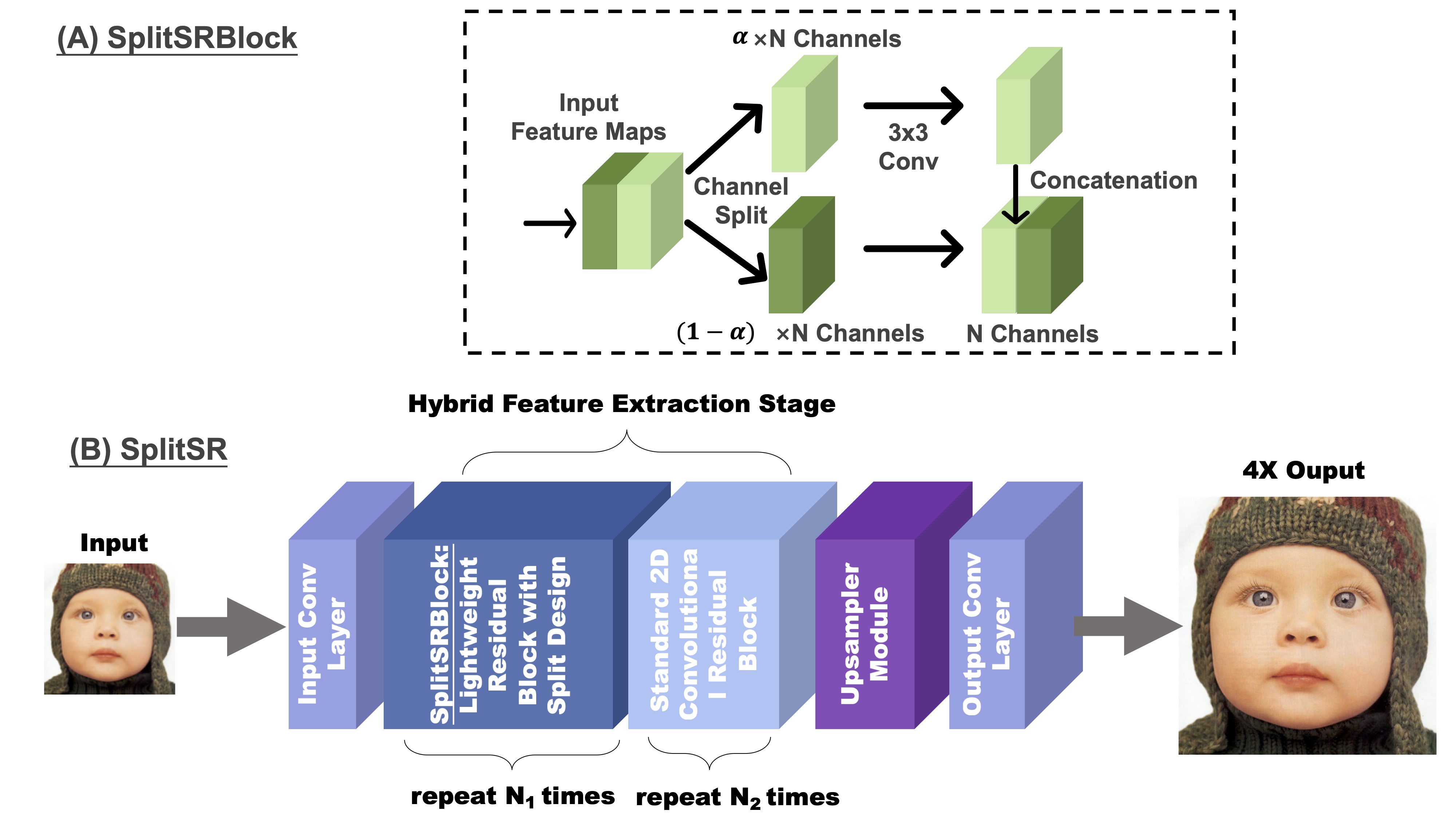}
  \caption{(A) \projectnameblock is a lightweight residual block based on standard 2D convolution to retain the quality of spatial transformation and reduces the computational cost by channel-split. (B) \projectname is a hybrid architecture that leverages a combination of lightweight residual blocks (e.g., \projectnameblock) and standard 2D convolutions so that people can optimize the tradeoff between latency and performance to enable SR on a constrained computational budget.}
  \label{fig:split_sr_all}
\end{figure*}

Existing lightweight residual blocks that leverage channel-splitting are currently designed for discriminative computer vision tasks like image classification. SR, however, is a generative task that usually requires significant fine-grained spatial information. When a small number of channels are involved in the computation after the channel-splitting stage, depthwise and pointwise convolutions sacrifice spatial information with only a small benefit in computational efficiency. Therefore, we propose the \projectnameblock (Figure~\ref{fig:split_sr_all}-A) specifically for efficient SR. Instead of depthwise and pointwise convolutions, the \projectnameblock uses standard 2D convolution after channel-splitting to retain spatial transformation. Standard convolutions preserve spatial information by having receptive fields, and the computation requirements for such operations are significantly diminished after channel-splitting.
As with the other lightweight residual blocks, the results of the two branches (processed and unprocessed) are concatenated at the end to retain the same number of output channels. However, the \projectnameblock reverses the direction of concatenation at the output so that the first $\alpha \times N$ channels (used for convolution) become the last $\alpha \times N$ channels of the output. By doing this, every single channel will be involved in a computation after $\frac{1}{\alpha}$ blocks. The theoretical computation reduction that can be obtained by using \projectname is $\alpha^{2}$, where $\alpha \in (0, 1]$. 

\subsection{\projectname Architecture Design}
\label{sec:method-hybrid}
Most modern SR networks share a similar four-stage structure \cite{lim2017enhanced, zhang2018image, dai2019second}: (1) an input layer like a single 2D convolutional layer, (2) a feature extraction stage with a large number of 2D convolutional layers to extract a high-dimensional representation of the image, (3) an upsampler stage made up of convolutional and pixel-shuffle layers to rearrange information along the depth dimension into blocks of spatial information, and (4) an output layer that generates the final upscaled image. Past attempts to make SR networks efficient enough to be run on mobile devices have involved replacing all of the standard convolution blocks with lightweight residual blocks \cite{lee2019mobisr}. Modern SR architectures usually have a large number of standard convolutional blocks, but the majority of them are clustered at the feature extraction stage. Each stage in a standard SR architecture serves a different purpose, so although na\"ively replacing standard convolution blocks at all locations increases efficiency, it often negatively impacts accuracy.

We introduce a hybrid architecture called \projectname to help balance the tradeoff between accuracy and latency for on-device SR (Figure~\ref{fig:split_sr_all}-B). \projectname is based on the RCAN model~\cite{zhang2018image}, although our hybrid approach could be applied to any other state-of-the-art SR model. RCAN uses 64 feature maps (output channels) in its convolutional layers and over 400 layers to achieve high accuracy. \projectname modifies RCAN by leveraging a combination of standard convolutional layers and lightweight residual blocks. Lightweight residual blocks with channel-splitting (e.g., \projectnameblock) can only be used at the feature extraction and upsampler stages since the input and output stages change the number of channels. As more lightweight residual blocks are introduced, \projectname becomes more efficient at the cost of sacrificing model complexity and degrees of freedom.
With this framing in mind, we introduce four parameters as potential design decisions for our proposed SR architecture:
\begin{enumerate}
    \item \textbf{Channel-Split Ratio:} the fraction of channels that are used for computation within each block
    \item \textbf{Hybrid Index:} the number of standard convolutional blocks that are replaced by lightweight residual blocks
    \item \textbf{Hybrid Mode:} the ordering of standard convolutional blocks and lightweight residual blocks \newline (e.g., standard-lightweight, lightweight-standard)
    \item \textbf{Replacement Location:} the stages where standard convolutional blocks are replaced \newline (e.g., feature extraction, upsampler)
\end{enumerate}

%% file: sections/experiments.tex
\section{Performance Evaluation \& Results}
\label{sec:experiment}
In this section, we describe the experiments we conducted to evaluate the effects of the various hyperparameters in our \projectname architecture. We also measure the performance benefits provided by our \projectnameblock for SR tasks compared to other block designs. To conclude this section, we compare \projectname against a state-of-the-art on-device SR model, MobiSR~\cite{lee2019mobisr}.

\subsection{Implementation}
\label{sec:experiment-implementation}

One of the challenges in deploying on-device deep learning models is the gap between high-level graph optimization and low-level operator optimization on the target platform. Unfortunately, very few hand optimization tools are designed to handle highly customized neural network architectures. Even after an architecture has been hand-optimized for a target device, moving that architecture to a new platform, or even changing input shapes, can require a complete rewrite of the optimized kernels. However, recent deep learning compilers like TVM~\cite{chen2018tvm} and Halide \cite{ragan2013halide} have emerged to simplify the optimization process. These compilers separate functional definitions from scheduling (i.e., how the code should be executed). This separation allows developers to quickly explore a schedule space without rewriting the kernel at each step.

In this work, we extended TVM to support the operations required for the \projectnameblock and the other three residual blocks described in the Section~\ref{sec:related-residual}. Our TVM-based on-device SR system directly converts a TensorFlow graph to a Relay representation \cite{roesch2018relay} and compiles that code to multiple target devices using LLVM. We utilize TVM's scheduling primitives for core operations in our proposed SR architecture (e.g., pixel shuffle). The scheduling primitives we apply are summarized as follow: 
\begin{itemize}
  \item \textbf{Tiling:} Splits the entire computation process into small blocks to better reuse data and utilize the cache. As a result, tiling improves computational efficiency and reduces memory traffic.
  \item \textbf{Packing:} Reformats the input tensor based on the tiled configuration to utilize memory more efficiently and reduce the cache misses.
  \item \textbf{Vectorization:} Takes advantage of single instruction, multiple data (SIMD) instructions in the target hardware to enable simultaneous multi-operator computation.
  \item \textbf{Unrolling:} Diminishes branch penalties and reduces latency while accessing memory and removing control flow operations.
\end{itemize}
Since we have implemented the operators following TVM's convention, we can utilize TVM's built-in graph optimization processes. These include dead code elimination, common sub-expression elimination, constant folding, and operator fusion. 

\subsection{Experimental Setup}
\label{sec:experiment-setup}
We deploy \projectname on an open-source embedded system called the Firefly-RK3399\footnote{\url{http://en.t-firefly.com/product/rk3399.html}}. The board is powered by dual 1.8~GHz clusters: one with two large Cortex-A72 cores and another with four small Cortex-A53 cores. The RK3399 also has a Mali-T860 MAP4 mobile GPU, but our work focus on evaluating the performance on CPU. By focusing on CPUs, the \projectname system can generalize to any ARM-based CPU mobile engine, which are common across contemporary mobile devices. Furthermore, GPUs can have unique architectures that are not available on other computing platforms (e.g., Raspberry Pi, low-end smartphones). We use single-precision floating-point arithmetic (FP32) as the precision of our models during evaluation.  

We implemented our system in TensorFlow~\cite{abadi2016tensorflow} and TVM~\cite{chen2018tvm}. During training, 96$\times$96 cropped high-resolution images and their low-resolution counterparts were pulled from the DIV2K dataset~\cite{timofte2017ntire} in batches of 16. Random horizontal flipping and random rotation at 90$^{\circ}$-intervals were used for data augmentation. The Adam optimizer~\cite{kingma2014adam} was used to minimize L1 loss with the following parameters: initial learning rate = $1\mathrm{e}{-4}$, $\beta_{1}$ = 0.9, $\beta_{2}$ = 0.999 and $\epsilon$ = $1\mathrm{e}{-7}$. Networks were trained for $6 \times 10^5$ steps with learning decay by a factor of 2 every $2 \times 10^5$ iterations.

In this work, we focus on models that increase resolution by $\times$4. To expedite model convergence, we first pre-train the model by training it to increase resolution by $\times$2 for $30 \times 10^5$ iterations. This training regiment has been widely used in past SR work~\cite{lee2019mobisr, zhang2018image, lim2017enhanced}.

\subsection{Evaluation Procedure and Metrics}
\label{sec:experiment-procedure}
To assess the performance of the models we deployed, we report both speed and accuracy as evaluation metrics. For speed, we ran each model 10 times and report averaged inference latency in milliseconds. For accuracy, we fixed random seed in training and report the standard metrics for SR image quality: peak signal-to-noise ratio (PSNR) and structural similarity (SSIM) \cite{wang2004image}. PSNR is computed on the Y channel of the YCbCr color space, which corresponds to image luminance. \textbf{PSNR is a logarithmic metric; hence, seemingly incremental differences are deceptively significant}. SSIM is a normalized metric (range: 0--1) for assessing the perceptual similarity between two images. 
We conduct our experiments on four datasets that are commonly used for SR model evaluations: Set5~\cite{bevilacqua2012low}, Set14~\cite{zeyde2010single}, B100~\cite{timofte2014a+}, and Urban100~\cite{huang2015single}. These datasets contain 5, 14, 100, and 100 images, respectively.

\subsection{Reference Model Configurations}
\label{sec:experiment-models}
We applied our hybrid design and \projectnameblock to the RCAN~\cite{zhang2018image} architecture, a state-of-the-art SR model. RCAN includes blocks of convolutional operations, which are clustered together to form residual groups with a short residual connection. By introducing channel-splitting, our approach enabled us to significantly extend depth of a SR network while maintaining the same inference latency. Because of RCAN's structure, which uses clusters of blocks as groups, we adapt our definition of hybrid index (HI) to refer to groups rather than blocks. We removed RCAN's channel attention layer because it is only beneficial when the number of channels in each convolution is high. Using our split design reduces the number of channels that are processed within each block to $\alpha \times N$, making the channel attention layer less necessary. Moreover, channel-splitting in the \projectnameblock provides some degree of attention to certain channels (a subset of feature maps that go through convolution).

Since \projectname is an architecture that can be tuned for computational budget, we created two versions of our model: 
\begin{itemize}
    \item \textbf{Accuracy-focused:} 7 residual groups each with 7 residual blocks
    \item \textbf{Latency-focused:} 5 residual groups each with 6 residual blocks
\end{itemize}
The accuracy-focused network configuration was meant to be directly comparable to MobiSR's reference model. The latency-focused model was more suitable for mobile applications given its shallowness and lower operation count.
We specifically designed the latency-focused configuration to have an average inference time around 500~ms, which we thought would demonstrate a better user experience and would be more accessible for low-end smartphones and more applicable to more real-world application. It is also worth noting that these model configurations represent two instantiations of the accuracy-latency tradeoff. The number of residual groups and residual blocks per group can be tuned depending on the computational budget and how much accuracy is valued over latency for a target application. The same can be applied for the number of feature maps (i.e., convolution channels) in the models.

We used MobiSR~\cite{lee2019mobisr} as a competitive benchmark since it also uses an RCAN-based design. \projectname is able to hold more residual blocks than MobiSR for the same computational budget because of the channel-splitting provided by the \projectnameblock. All of the models in our experiments had 16 feature maps to remove that hyperparameter as a possible confound.

\subsection{Hyperparameter Selection for \projectname}
\label{sec:experiment-hyperparams}
\projectname has many hyperparameters that can be adjusted to balance accuracy and performance. 
Below, we describe the effects of some of these hyperparameters on our latency-focused model. 

\subsubsection{Channel-Split Ratio}
Table~\ref{table:alpha} shows the performance implications of different values for the channel-split ratio $\alpha$. As $\alpha$ increases, more channels are used for computation. This creates a more complex model that is able to generate a higher-order representation of the data, leading to higher accuracy. However, increasing $\alpha$ also leads to higher latency. The latency when $\alpha=0.25$ is very close to the latency when $\alpha=0.125$ while yielding better PSNR for all datasets. Although $\alpha=0.5$ and $\alpha=1$ provide better accuracy, they take much longer for inference. To establish a good trade-off between latency and accuracy, we use $\alpha$=0.25 for remainder of our evaluation.

\begin{table*}[t!]
\centering
\begin{tabular}{c c c | c c c c|} \cline{4-7}
\multicolumn{3}{l|}{}  & \multicolumn{4}{c|}{\textbf{Average PSNR (dB) / SSIM}} \\ \hline
\multicolumn{1}{|c}{\textbf{Alpha}} & \textbf{Params} & \textbf{Latency (ms)} & \textbf{Set5} & \textbf{Set14} & \textbf{B100} & \textbf{Urban100} \\ \hline

\multicolumn{1}{|c}{$\alpha$ = 0.125} & 90k & 603 & 31.46 / 0.8942 & 28.12 / 0.7876 & 27.25 / 0.7451 & 25.14 / 0.7686 \\ 
 
\multicolumn{1}{|c}{$\alpha$ = 0.250} & 94k & 629 & 31.53 / 0.8950 & 28.18 / 0.7887 & 27.28 / 0.7458 & 25.20 / 0.7704 \\ 
 
\multicolumn{1}{|c}{$\alpha$ = 0.500} & 110k & 725 & 31.57 / 0.8958 & 28.20 / 0.7897 & 27.29 / 0.7467 & 25.24 / 0.7724 \\ 
 
\multicolumn{1}{|c}{$\alpha$ = 1.000} & 172k & 856 & 31.75 / 0.8980 & 28.30 / 0.7918 & 27.36 / 0.7486 & 25.41 / 0.7787 \\ \hline
\end{tabular}
\caption{The performance comparison for different settings for the channel-split ratio $\alpha$ on our latency-focused \projectname model (HI = 3, hybrid mode = front, replacement location = feature extraction).}
\label{table:alpha}
\end{table*}

\begin{table*}[t!]
\small
\centering
\begin{tabular}{|c c c | c c c c|} \cline{4-7}
\multicolumn{3}{l|}{} & \multicolumn{4}{c|}{\textbf{Average PSNR (dB) / SSIM}} \\ \hline
\multicolumn{1}{|c}{\textbf{Hybrid Index (HI)}} & \textbf{Params} & \textbf{Latency (ms)} & \textbf{Set5} & \textbf{Set14} & \textbf{B100} & \textbf{Urban100} \\ \hline

\multicolumn{1}{|c}{HI = 2} & 120k & 706 & 31.61 / 0.8960 & 28.22 / 0.7900 & 27.31 / 0.7472 & 25.28 / 0.7739 \\ 
 
\multicolumn{1}{|c}{HI = 3} & 94k & 629 & 31.53 / 0.8950 & 28.18 / 0.7887 & 27.28 / 0.7458 & 25.20 / 0.7704 \\ 
 
\multicolumn{1}{|c}{HI = 4} & 68k & 555 & 31.33 / 0.8920 & 28.05 / 0.7856 & 27.19 / 0.7432 & 25.02 / 0.7633\\ 
\hline
\end{tabular}
\caption{The performance comparison for different settings for the hybrid index HI on our latency-focused \projectname model ($\alpha$ = 0.25, hybrid mode = front, replacement location = feature extraction).}
\label{table:hybrid_index}
\end{table*}

\begin{table*}[t!]
\centering
\begin{tabular}{|c c c | c c c c|} \cline{4-7}
\multicolumn{3}{l|}{}  & \multicolumn{4}{c|}{\textbf{Average PSNR (dB) / SSIM}} \\ \hline
\multicolumn{1}{|c}{\textbf{Hybrid Mode}} & \textbf{Params} & \textbf{Latency (ms)} & \textbf{Set5} & \textbf{Set14} & \textbf{B100} & \textbf{Urban100} \\ \hline

\multicolumn{1}{|c}{Front} & 94k & 629 & 31.53 / 0.8950 & 28.18 / 0.7887 & 27.28 / 0.7458 & 25.20 / 0.7704 \\ 

\multicolumn{1}{|c}{End} & 94k & 629 & 31.47 / 0.8943 & 28.12 / 0.7876 & 27.25 / 0.7451 & 25.15 / 0.7688 \\ 

\multicolumn{1}{|c}{Mixed} & 94k & 629 & 31.46 / 0.8945 & 28.14 / 0.7880 & 27.25 / 0.7450 & 25.15 / 0.7688 \\ \hline
\end{tabular}
\caption{The performance comparison for three different hybrid modes --- front, end, and mixed --- on our latency-focused \projectname model ($\alpha$ = 0.25, HI = 3, replacement location = feature extraction).}
\label{table:hybrid_mode}
\end{table*}

\begin{table*}[t!]
\small
\centering
\begin{tabular}{|c c c | c c c c|} \cline{4-7}
\multicolumn{3}{l|}{} & \multicolumn{4}{c|}{\textbf{Average PSNR (dB) / SSIM}} \\ \hline
\multicolumn{1}{|c}{\textbf{Replacement Location}} & \textbf{Params} & \textbf{Latency (ms)} & \textbf{Set5} & \textbf{Set14} & \textbf{B100} & \textbf{Urban100} \\ \hline

\multicolumn{1}{|c}{Feature Extraction (FE)} & 94k & 629 & 31.53 / 0.8950 & 28.18 / 0.7887 & 27.28 / 0.7458 & 25.20 / 0.7704 \\ 

\multicolumn{1}{|c}{FE + Upsampling} & 80k & 504 & 31.45 / 0.8937 & 28.11 / 0.7874 & 27.24 / 0.7447 & 25.11 / 0.7675 \\ 

\multicolumn{1}{|c}{Throughout} & 67k & 468 & 31.10 /0.8877 & 27.91 / 0.7827 & 27.10 / 0.7403 & 24.84 / 0.7563 \\ \hline
\end{tabular}
\caption{The performance comparison for different lightweight block placements on our latency-focused \projectname model ($\alpha$ = 0.25, HI = 3, hybrid mode = front).}
\label{table:block_location}
\end{table*}

\subsubsection{Hybrid Index}
Table~\ref{table:hybrid_index} shows the performance of different values for the hybrid index (HI). When HI increases, more standard 2D convolution groups are replaced with lightweight residual groups, resulting in both lower latency and accuracy; nevertheless, mixing standard and lightweight convolutions can lead to a sweet spot for accuracy and latency. The accuracy gap between HI=3 and HI=4 is substantially larger than that between HI=2 and HI=3. Meanwhile, the latency difference between those pairs is practically equivalent. For these reasons, we use HI=3 for future evaluation.  

\subsubsection{Hybrid Mode}
Table~\ref{table:hybrid_mode} shows the performance implications of different hybrid modes. We examined three different hybrid modes: 
\begin{enumerate}
    \item \textbf{Front:} The first HI residual groups are replaced
    \item \textbf{End:} The last HI residual groups are replaced
    \item \textbf{Mixed:} HI residual groups are replaced with a regular spacing
\end{enumerate}
Because HI is fixed in this experiment (HI=3), the model complexity remains the same; only the order of operations changes across the different hybrid modes. Therefore, the inference latency is identical across all three hybrid modes. The ``front'' configuration outperforms the other two modes in regards to accuracy, so we use that setting for future evaluation. 

\begin{figure*}[t!]
  \includegraphics[width=10cm]{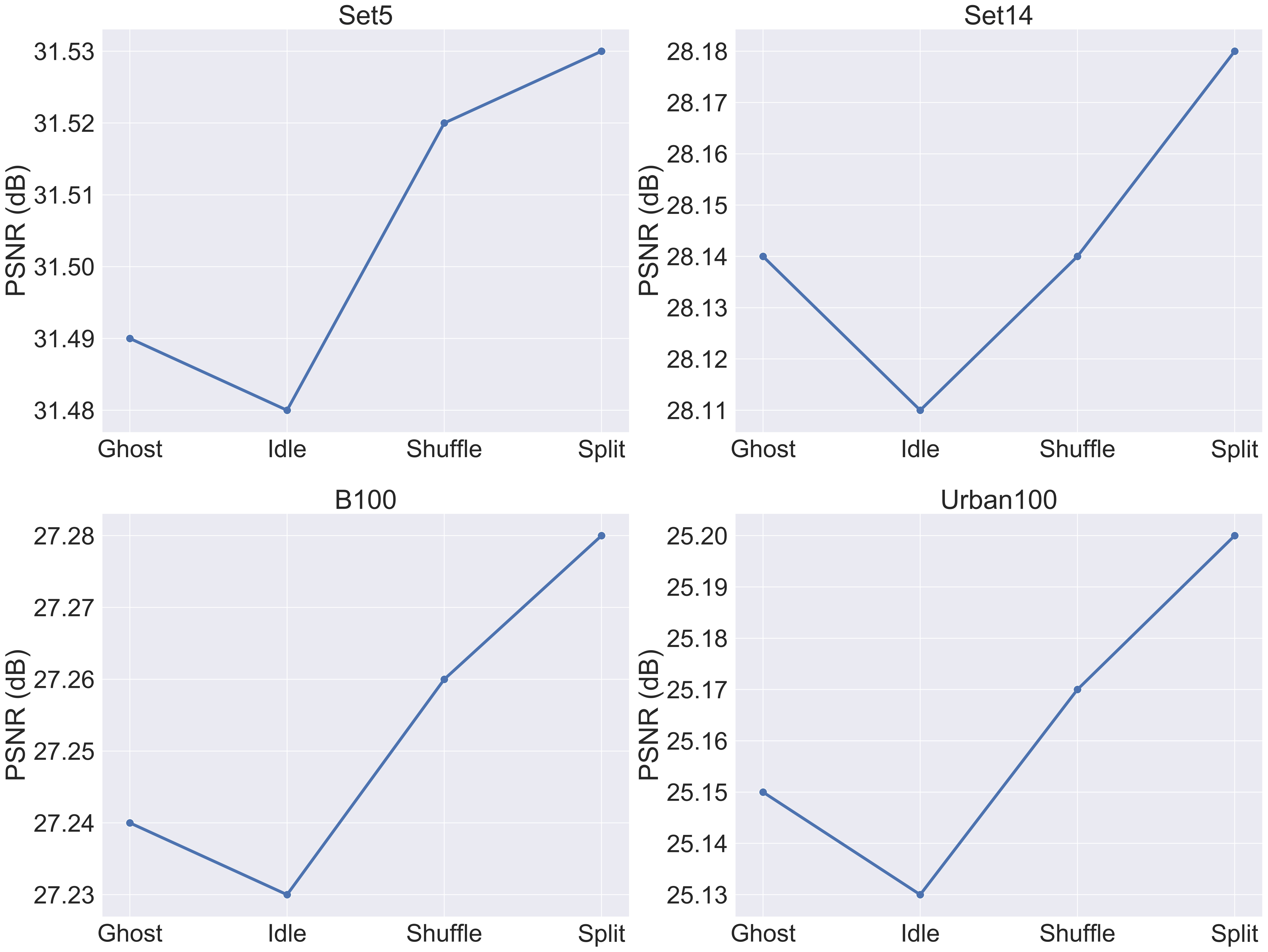}
  \centering
  \caption{The effect that residual block selection has on PSNR for (top-left) Set5~\cite{bevilacqua2012low}, (top-right) Set14~\cite{zeyde2010single}, (bottom-left) B100~\cite{timofte2014a+}, (bottom-right) and Urban100~\cite{huang2015single}.}
  \label{figure:blocks}
\end{figure*}

\subsubsection{Replacement Location}
Table~\ref{table:block_location} shows the performance implications of where the lightweight residual blocks are used to replace standard convolutional blocks within the RCAN architecture. As discussed in the Section~\ref{sec:method-hybrid}, the two major stages where it makes the most sense to utilize lightweight residual blocks are the feature extraction and upsampling stages. However, standard convolutional blocks can also be replaced at the output of each residual group and at the output of the feature extraction stage. We investigate the following replacement strategies:
\begin{enumerate}
    \item \textbf{Feature extraction only:} Convolutions are replaced only in the feature extraction stage
    \item \textbf{Feature extraction + upsampling:} Convolutions are replaced in both the feature extraction and upsampling stages
    \item \textbf{Throughout:} Convolutions are replaced in the feature extraction stage, the upsampling stage, at the end of each residual group, and the output of the feature extraction stage
\end{enumerate}
Additional strategies could have included replacing just the convolutions in the upsampling stage or the other non-major stages, but those stages are much smaller than the main feature extraction stage and would therefore not significantly impact the network's performance. We find that that replacing convolutions outside of the feature extraction stage deteriorates accuracy without commensurate latency benefits, so we only replace convolutions in the feature extraction stage for future investigation. 

\subsubsection{Lightweight Residual Block Type}
\begin{figure}[t!]
  \includegraphics[width=10cm]{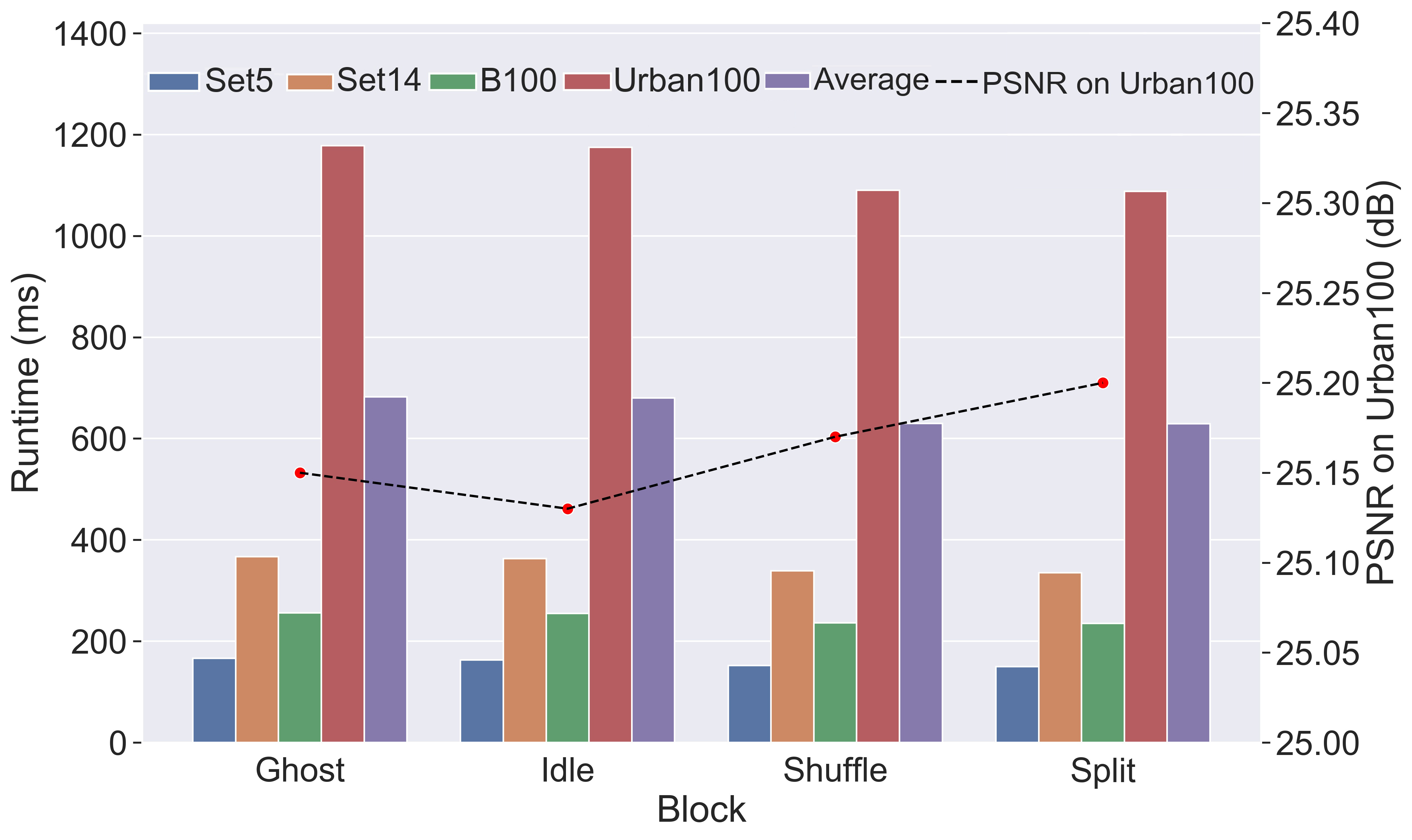}
  \centering
  \caption{The latency and accuracy of our latency-focused \projectname model across four datasets: Set5~\cite{bevilacqua2012low}, Set14~\cite{zeyde2010single}, B100~\cite{timofte2014a+}, and Urban100~\cite{huang2015single}.}
  \label{figure:latency_focused_model}
\end{figure}

We compare the performance engendered by four lightweight residual block designs: he ShuffleBlock~\cite{ma2018shufflenet}, the IdleBlock~\cite{xu2019hybrid}, the GhostBlock~\cite{han2019ghostnet}, and our proposed \projectnameblock.
Figure~\ref{figure:latency_focused_model} illustrates the performance RCAN achieves with the various lightweight residual blocks for the four validation datasets. \projectnameblock outperforms the other blocks in terms of both latency and accuracy. The ShuffleBlock is the clear runner-up in regards to both metrics. The IdleBlock results in the worst accuracy, whereas the GhostBlock produces the worst latency. Both the IdleBlock and the GhostBlock expand feature maps close to their input, whereas the ShuffleBlcok and \projectnameblock do not. Our empirical results show that expansion does not lend to improved SR performance.

Figure~\ref{figure:blocks} expands on the accuracy of the various models across the validation datasets. Our proposed \projectnameblock surpasses the other three residual blocks across all the datasets, and each dataset exhibits the same ranking for the different blocks. These findings validate our hypothesis that utilizing the split design with conventional 2D convolution achieve a superior balance in accuracy and speed.

\begin{table*}[t!]
\centering
\small
\begin{tabular}{c c c | c c c c|} \cline{4-7}
\multicolumn{3}{l|}{}  & \multicolumn{4}{c|}{\textbf{Average PSNR(dB) / SSIM}} \\ \hline
\multicolumn{1}{|c}{\textbf{Model}} & \textbf{Params} & \textbf{Latency (ms)} & \textbf{Set5} & \textbf{Set14} & \textbf{B100} & \textbf{Urban100} \\ \hline

\multicolumn{1}{|c}{MobiSR} & 152k & 4570 & 31.73 / 0.8873 & 28.24 / 0.7729 & 27.33 / 0.7283 & 25.34 / 0.7610\\ 

\multicolumn{1}{|c}{MobiSR (w/ shuffle)} & 17k & 1023 & N/A & N/A & N/A & 24.57 / N/A \\ 

\multicolumn{1}{|c}{MobiSR (dual-model)} & 182k & 1426* & 31.40 / N/A & 28.10 / N/A & 27.30 / N/A & 25.30 / N/A \\

\multicolumn{1}{|c}{Bilinear} & N/A & N/A & 27.56 / 0.80 & 25.51 / 0.69 & 25.54 / 0.66 & 22.69 / 0.65 \\ \hline

\multicolumn{1}{|c}{SplitSR (accuracy)} & 174k & 947 & 31.76 / 0.8982 & 28.29 / 0.7916 & 27.39 / 0.7491 & 25.46 / 0.7795 \\ 

\multicolumn{1}{|c}{SplitSR (latency)} & 94k & 629 & 31.53 / 0.8950 & 28.18 / 0.7887 & 27.28 / 0.7458 & 25.20 / 0.7704 \\ \hline
\end{tabular}
\\\small{*This experiment required a DSP and a CPU.}
\caption{The comparison between the results reported in the MobiSR paper~\cite{lee2019mobisr} and \projectname deployed on TVM~\cite{chen2018tvm}.}

\label{table:mobisr_vs_splitsr}
\end{table*}

\subsection{Overall Performance of \projectname}
\label{sec:experiment-overall}
We also compare \projectname against the current state-of-the-art mobile SR system, MobiSR~\cite{lee2019mobisr}. In that work, Lee et al. evaluate three different configurations:
\begin{enumerate}
    \item \textbf{MobiSR:} This is the most standard network with normal convolutions, but fewer layers so that their model can operate on a mobile device
    \item \textbf{MobiSR w/ Shuffle:} In this model, standard convolutional blocks are replaced with ShuffleBlocks since those were determined to be the best performing lightweight residual block in Lee et al.'s experiments.
    \item \textbf{MobiSR Dual-Model:} In this model, MobiSR is split into two models that are run on heterogeneous mobile processors, integrating the CPU, GPU, and digital signal processors (DSP). A heavy model is run on the DSP while a lightweight model is run on the CPU/GPU. To utilize these models, MobiSR splits a raw image into small patches (90$\times$160) and feeds them into an load-balancing unit that assesses their difficulty; `hard' patches are sent to the DSP and `easy' patches are sent to the CPU/GPU for SR inference.
\end{enumerate} 
To test the performance of MobiSR, Lee et al. used a DSP and an Octa-core CPU with four large customized ARM A75 cores and four A55 cores with a clock speed up to 2.8GHz (Qualcomm Snapdragon 845 SoC\footnote{\url{https://www.qualcomm.com/products/snapdragon-845-mobile-platform}}). In contrast, we evaluate \projectname on a CPU with four A72 cores and two A53 cores; the clock speed of the cores can go only go up to 1.8~GHz, comparable to the much older Qualcomm Snapdragon 650\footnote{\url{https://www.qualcomm.com/products/snapdragon-650-mobile-platform}}.

Table~\ref{table:mobisr_vs_splitsr} shows the performance of two \projectname configurations (latency- and accuracy-focused) against the aforementioned MobiSR configurations. We also include the performance of bilinear interpolation, which is the existing standard for on-device SR. Note that Lee et al. did not evaluate all three of their configurations on all possible datasets; thus leaving gaps in our comparison. We found that \projectname is 4.78$\times$ faster and achieves higher accuracy on every validation dataset compared to the standard MobiSR model using single processor. When we compare our latency-focused model against MobiSR with shuffle, we demonstrate superior accuracy in the Urban100 dataset as well as 33\% acceleration in CPU latency. When we compare \projectname against MobiSR's dual-model configuration, our accuracy-based model not only achieves higher accuracy but also a $\sim$34\% speedup.

In summary, \projectname achieves better accuracy and an inference speedup of up to 4.8$\times$ despite the fact that MobiSR uses higher performance hardware and an intricate system to maximize performance. By leveraging a compiler dedicated to deep learning for generative vision tasks like SR, our results  suggest that people should integrate lightweight residual blocks along with a deep learning compiler into their heavy mobile computing system to gain substantial speedup. Moreover, our proposed \projectnameblock exhibits superior performance against alternatives proposed by prior work, showing that depthwise separable convolution is not always necessary for new lightweight residual blocks. 

%% file: sections/user_study.tex
\section{User Study}
\label{sec:userstudy}
To contextualize the results of our offline evaluation, we took a step further to deploy \projectname on a smartphone and investigate whether individuals can notice the difference in image quality provided by \projectname over the existing standard of bilinear interpolation.
We first describe the way we implemented \projectname as part of a standard image gallery app called \appname to provide as seamless of an experience as possible.
We then describe the two tasks that participants were asked to perform and the associated findings.

\subsection{\appname Design}
\label{sec:userstudy-app}
For our study, we developed \appname as an image gallery app for viewing images saved on the smartphone. We deployed this app on a Google Pixel 4, which has a 5.7-inch OLED touch screen with a resolution of 2280$\times$1080 and a Qualcomm Snapdragon 855 processor. Images that are loaded in \appname are displayed in a 1742$\times$1080-pixel area, with the rest of the screen being reserved for instructions and interface controls. Whenever the user loads an image onto the screen, \appname randomly selects either \projectname or bilinear interpolation for upsampling during zoom. Using a button at the bottom of the screen, the user can see the same image post-processed by the alternate SR method.

Most images that people view on their smartphones nowadays have a larger resolution than the images in the datasets that we used to evaluate our technique, and processing larger images incurs longer delays. Bilinear interpolation is able to process such images with relative ease, but we had to intelligently utilize \projectname to achieve tolerable processing times. Within \appname, we use the latency-focused \projectname model to upsample 256$\times$256~px patches to 4$\times$. If the user's desired zoom level is below 2$\times$, \appname simply uses bilinear upsampling to process the patches. If the desired zoom level is between 2--4$\times$, \appname pushes patches through the model to upsampling them by a factor of 4$\times$ and then downsamples the results to the desired resolution. If the desired zoom level is greater than 4$\times$, \appname still pushes the patches through the \projectname model, but then applies bilinear interpolation to the results to upsample further. We limit the maximum zoom level to 5$\times$ for our study. We prioritize processing the patches closest to where the user initiates their pinch-to-zoom gesture to ensure that the most important parts of the image are done first. Patches outside of the user's final view of the image are still processed since users should still be allowed to explore the rest of the image during or after a zoom gesture.

\begin{figure*}[t!]
  \includegraphics[width=\textwidth]{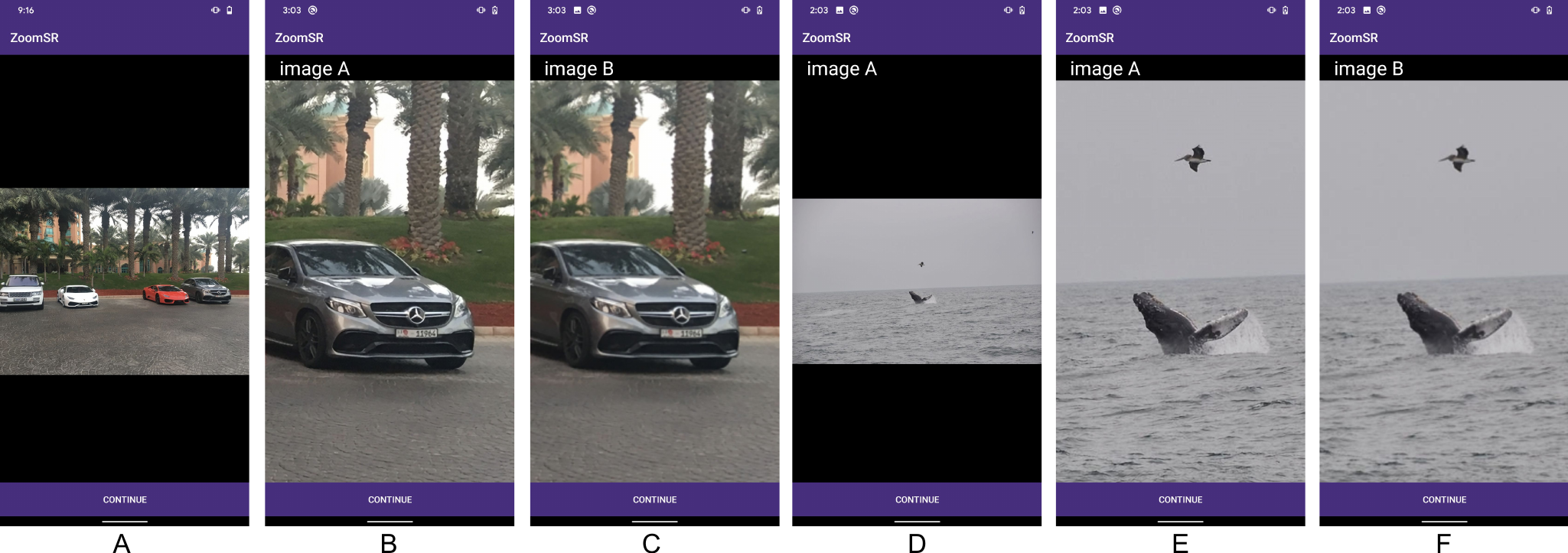}
  \caption{Screenshots taken from \appname for the image task: (A--C) an image of cars and (D--F) an image of animals. The screenshots show the (A, D) original images, (B, E) images after upsampling with \projectname, and (C, F) images after upsampling with bilinear interpolation. Note that the presentation order of \projectname and bilinear interpolation was randomized to mitigate ordering effects.}
  \label{fig:demo_zoomsr_image}
\end{figure*}

\subsection{Protocol}
\label{sec:userstudy-protocol}
We conducted a user study with 15 participants (8 male, 7 female) to assess the users' perception of the image quality engendered by \projectname. The study was split into two parts --- an image task and a reading task --- since people can fixate on different characteristics of objects and text, respectively. 

For the image task, participants were asked to sequentially examine 10 randomly ordered images of subjects like animals, people, and nature (see Figure \ref{fig:demo_zoomsr_image}). Participants were free to pan and zoom the image as they pleased. When they were done examining the same image using the two different SR methods --- \projectname and bilinear interpolation --- participants were asked to rate the quality of each version of the image. Participants were asked to rate the images according to how they were rendered (e.g., resolution, focus, blurriness) rather than their content. The ratings were made along a 7-point Likert scale (1: lowest quality, 7: highest quality).

\begin{figure*}[t!]
  \includegraphics[width=13cm]{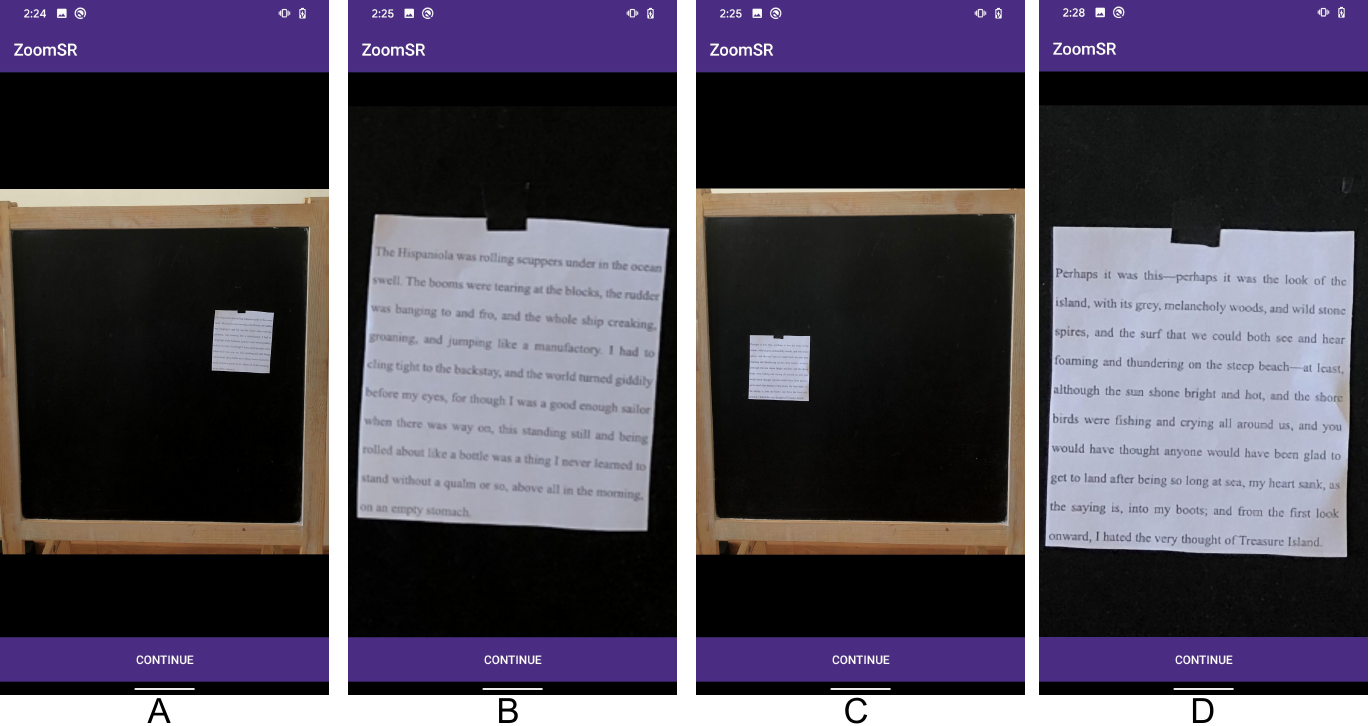}
  \caption{Screenshots taken from \appname for the reading task. The screenshots show the (A, C) original images, (B) an image after upsampling with bilinear interpolation, and (D) an image after upsampling with \projectname. Note that the presentation order of \projectname and bilinear interpolation was randomized to mitigate ordering effects.}
  \label{fig:demo_zoomsr_Text}
\end{figure*}

For the reading task, participants were asked to sequentially examine 10 images of text that was photographed on a distant blackboard (see Figure \ref{fig:demo_zoomsr_Text}). The blackboard was 2 meters away and the text had roughly 60--70 characters, so participants had to zoom-in to resolve the characters. The text was also placed at a different location on the blackboard each time to prevent participants from learning the same zooming pattern over time. Participants were asked to read text and to rate the ease with which they were able to do it in along a 7-point Likert scale (1: very easy, 7: very hard). Unlike with the previous task, participants only saw each image with either \projectname or bilinear interpolation (evenly distributed) since seeing the same image twice would have made the latter far easier to read.

\begin{figure*}[t!]
  \includegraphics[width=\textwidth]{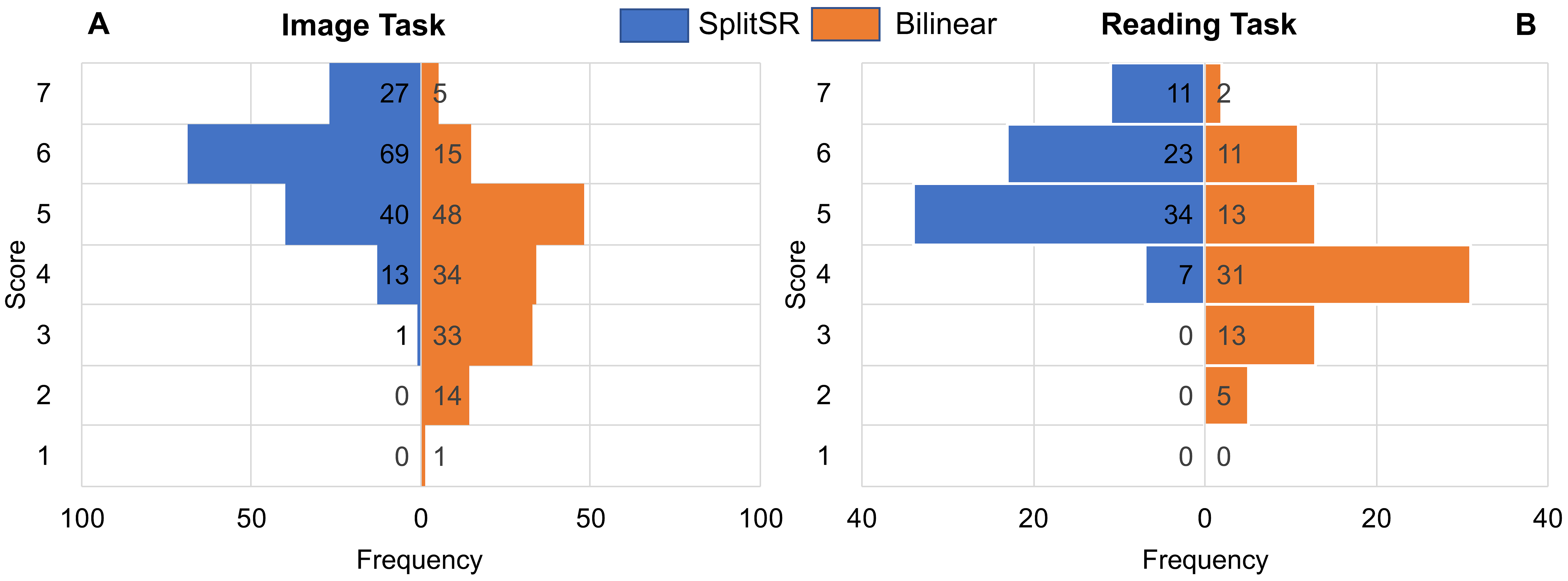}
  \centering
  \caption{The distributions of Likert scores in the (left) image and (right) reading tasks.}
  \label{figure:zoom_sr_scores}
\end{figure*}

\subsection{Results}
Participants in both studies did not report noticing the additional latency incurred by \projectname versus bilinear interpolation. Pinch-to-zoom gestures take roughly 1000~ms on average, and the processing time of the latency-focused \projectname configuration is takes 629~ms on average. \appname starts upsampling images as soon as a gesture is initiated, which means that \projectname typically finished processing images before people completed their gestures and returned to examining the images' details.

\label{sec:userstudy-results}
\subsubsection{Part 1: Image Task}
Figure \ref{figure:zoom_sr_scores}A summarizes the distribution of Likert scores for the image task. A statistically significant difference was found between the Likert scores for bilinear interpolation and \projectname ($Z=-9.270, p<0.01$) according to the Wilcoxon signed-rank test. The median score earned by \projectname was 2 points higher than that of bilinear interpolation (6 vs. 4). This result was upheld in a random effects model that accounted for the image ID, participant ID, and presentation order of the different conditions ($F=285, p<0.01$). Across the 150 pairwise comparisons (15 participants $\times$ 10 images), participants perceived that \projectname improved image quality over bilinear interpolation for 85.3\% of the images. Participants only preferred the images from bilinear interpolation 4.7\% of the time, and the remaining cases has no perceptible difference according to participants.

\subsubsection{Part 2: Reading Task}
For this analysis, we use the Mann-Whitney U test to compare Likert scores since the measurements were unpaired. Figure \ref{figure:zoom_sr_scores}B demostrates the distribution of Likert score of the reading task. As before, a statistically significant difference was found in favor of \projectname ($Z=-6.486, p<0.01$). The median Likert score for \projectname was 1 points higher than that of bilinear interpolation (5 vs. 4). This results was upheld in a random effects model that accounted for the image ID, participant ID, and presentation order of the different conditions ($F=155, p<0.01$).

%% file: sections/conclusion.tex
\section{Discussion \& Limitations}
\label{sec:discussion}
In this work, we demonstrate the first instance of deep learning-based SR on a mobile device using our \projectname architecture. The channel-split design of our \projectnameblock made this goal tenable by yielding superior accuracy over prior approaches while achieving faster inference speed. These innovations not only made it possible to perform on-device SR, but also improved image quality by a noticable amount for end-users. In this section, we describe potential avenues of future work and the limitations of our findings.

\begin{table*}[t!]
\small
\centering
\begin{tabular}{c c c | c c c c|} \cline{4-7}
\multicolumn{3}{l|}{}  & \multicolumn{4}{c|}{\textbf{Average PSNR(dB) / SSIM}} \\ \hline
\multicolumn{1}{|c}{\textbf{Model}} & \textbf{Params} & \textbf{Latency (ms)} & \textbf{Set5} & \textbf{Set14} & \textbf{B100} & \textbf{Urban100} \\ \hline

\multicolumn{1}{|c}{MobiSR (w/ TVM)} & 172K & 941 & 31.73/0.8976 & 28.25/0.7908 & 27.33/0.7481 & 25.39/0.7780\\ 

\multicolumn{1}{|c}{\textbf{\projectname (accuracy)}} & 174K & 947 & \textbf{31.76/0.8982} & \textbf{28.29/0.7916} & \textbf{27.39/0.7491} & \textbf{25.46/0.7795}\\ \hline

\multicolumn{1}{|c}{Light MobiSR (w/ TVM)} & 101K & 626 & 31.38/0.8923 & 28.07/0.7860 & 27.21/0.7439 & 25.07/0.7655\\ 
 
\multicolumn{1}{|c}{\textbf{\projectname (latency)}} & 94K & 629 & \textbf{31.53/0.8950} & \textbf{28.18/0.7887} & \textbf{27.28/0.7458} & \textbf{25.20/0.7704}\\ 

 \hline
\end{tabular}
\caption{The performance comparison between MobiSR and \projectname when both are implemented with TVM~\cite{chen2018tvm}.}
\label{tab:tvm_mobisr_splitsr}
\end{table*}

\subsection{Importance of TVM}
\label{sec:discussion-tvm}
As noted in Section~\ref{sec:experiment-implementation}, one of the challenges in deploying deep learning models onto platforms like mobile devices is the gap between high-level graph optimization and low-level operator optimization on the target platform. TVM addresses this issue by automatically generating efficient machine-readable code for deep learning models, which is why we use it to develop \projectname and \appname. In order to quantify the performance improvement that can be attributed to our integration with TVM, we also deployed \projectname using the desktop-based Tensorflow without TVM. Note that the desktop-based TensorFlow is not designed for ARM-based platforms, and we were unable to use TensorFlow Lite since it does not support some of the operations we needed to implement \projectname. The accuracy-focused configuration of \projectname takes an average of 2560~ms for inference across all the benchmark datasets and the latency-focused configuration takes 1802~ms, highlighting that optimization was critical for on-device deployment. These results do not invalidate the results presented earlier since prior literature has also leveraged deep learning optimization to achieve their results. Our performance baseline, MobiSR~\cite{lee2019mobisr}, used the Snapdragon Neural Processing Engine to optimize operations on their deep learning architecture. MobiSR also uses the Snapdragon 845, which has superior hardware specifications to the Firefly-RK3399 we used in our evaluations.

We felt that comparing \projectname with TVM against the results presented in the MobiSR paper (with its own optimization) was the most honest comparison that could be made since MobiSR's source code was not available and we were unable to reproduce their results. Nevertheless, we conducted an additional evaluation comparing \projectname and MobiSR with standardized hardware and TVM as the compiler (Table~\ref{tab:tvm_mobisr_splitsr}). To produce comparable results to those shown in the MobiSR paper, we deployed their single-model architecture with a non-reduced upsampler module, adding an additional 20k parameters out of 170k. Comparing this model to the accuracy-focused \projectname model, we find that \projectname outperforms MobiSR by 0.05 dB on average across all four validation datasets while achieving similar inference latency. We also compare the latency-focused \projectname model against a lightweight version of MobiSR with 6 residual blocks in each group instead of 10 residual blocks. We find that \projectname outperforms MobiSR in this evaluation as well, providing a significant accuracy boost of 0.13 dB on the Urban100 dataset.

\subsection{Future Applications of \projectname}
\label{sec:discussion-future}
In this paper, we based our model architecture on RCAN~\cite{zhang2018image}, but our hybrid architecture and use of channel-splitting can be extended to other modern SR architectures. We also hypothesize there may be an opportunity to apply \projectnameblock to the comparable task of on-device audio SR~\cite{kuleshov2017audio}, which increases the resolution of an audio spectrogram to achieve higher quality audio. In general, we encourage researchers to explore new ways of integrating our \projectnameblock in places where they would normally use standard convolution to support on-device computation on different hardware platforms. 

We foresee many potential applications that could benefit from on-device SR beyond our \appname app. Studies have shown that SR can play a major role in medical imaging~\cite{robinson2010new, greenspan2009super}. As medical devices become more ubiquitous, particularly in low- and middle-income regions, the ability to run clinical assessments without requiring a data connection will become increasingly important. Mobile health would also benefit from SR. For example, Liu et al.~\cite{liu2020multi} recently demonstrated that smartphones can be used for remote physiological testing (i.e., photoplethysmography), and McDuff~\cite{mcduff2018deep} showed that SR improves the accuracy of such techniques. Another application that would benefit from SR is PupilScreen~\cite{Mariakakis2016}, a screening tool for traumatic brain injuries that utilizes a smartphone's camera measure the pupillary light reflex. Because the camera is being used to measure the size of a person's pupils, the video resolution has a direct impact on PupilScreen's precision and diagnostic accuracy. Although computations for both of these applications could be done on the cloud, bypassing the need for data upload would help maintain the privacy of protected health information.

\subsection{Pinch-to-Zoom Gesture Prediction for Lower Latency}
\label{sec:discussion-gesture}
Regardless of a model's performance offline, the way the model is integrated into the user experience has a significant impact on latency (and therefore, perceived performance). For our study app, we used a single \projectname model to upsample an image by 4$\times$ and then adjusted the image based on the final zoom level. This decision was a compromise that had consequences for images that needed to be upsampled at both lower and higher zoom levels. For images that had to be upsampled past 4$\times$, we used a smaller model with quicker latency to ensure that \appname was responsive, but that led to sacrifices in image quality. For images that did not have to be upsampled that high, we achieve the maximum image quality possible, but the model was larger than necessary and thus \appname spent more time processing the image than it would have required had it used a smaller model. These drawbacks can be mitigated by simultaneously running a small set of models that upsample images to varying scales (i.e., 2$\times$, 3$\times$, and 4$\times$). However, this is problematic to do even with our state-of-the-art latency, and the memory footprint of these models also imposes its own challenges. We believe that the final zoom level can be anticipated in some cases by observing the way a person performs a pinch-to-zoom gesture on the touchscreen, with larger gestures indicating higher upsampling. Knowing this information, a scheduler can prioritize the model with the closest scaling factor to minimize the amount of post-processing required. Investigating such a prediction mechanism and determining the ideal number of models to optimize latency remain as topics for future investigation.

\subsection{Limitations}
\label{sec:discussion-limits}
We investigated five hyperparameters in the \projectname system: channel split ratio ($\alpha$), hybrid index, hybrid mode, replacement location, and block type. In our experiments, we varied a single hyperparameter while keeping the others constant, thus allowing us to report the effect of each parameter on their own. We hypothesize that a complete grid search across all possible hyperparameters may lead to a more optimal configuration and improved performance, but we suspect that the configuration we presented for our \projectname system is close to that target. AutoTVM~\cite{chen2018learning}, which is made by the same researchers who created TVM, provides the additional functionality of supporting automatic hyperparameter search, which we anticipate would be particularly useful for identifying the optimal patch scheduling primitives (e.g., patch size) to attain even lower latency.

Other limitations of our evaluation lie in the fact that we only tested one deployment configuration for \projectname: an image gallery app on a Google Pixel 4 with a single model for upsampling images by 4$\times$. We chose to deploy \appname on Android over iOS since TVM is more compatible with Android, and we chose the zoom factor of 4$\times$ to have the fairest comparison possible with Lee et al.'s MobiSR work~\cite{lee2019mobisr}. Nevertheless, \projectname can be used on a variety of platforms with different configurations, and we leave this exploration to future work.

\section{Conclusion}
\label{sec:conclusion}
In this paper, we have proposed a novel and end-to-end mobile super-resolution system called \projectname. By introducing a split design into modern SR architectures and leveraging a modern deep learning compiler to perform low-level operator optimization, we have achieved up to 5$\times$ inference speedup while outperforming accuracy compare against the previous state-of-the-art systems. We also developed and tested \appname, the first-ever smartphone app for on-device SR using deep learning. We believe our proposed approach will enable many future on-device SR applications, particularly healthcare applications in low-resource settings where cloud connectivity and on-device compute capacity is limited. It is our hope that other researchers apply our findings to this domain and others in the future.